\documentclass[11pt]{article}
\usepackage[T1]{fontenc}
\usepackage{amssymb}
\usepackage{amsfonts}
\usepackage{graphicx}
\usepackage{amsmath}

\usepackage{hyperref}

\makeatletter \@addtoreset{equation}{section} \makeatother

\newtheorem{theorem}{Theorem}

\newtheorem{remark}{Remark}

\newtheorem{proposition}{Proposition}

\def\la{\langle}
\def\ra{\rangle}
\begin{document}

\title{Quantum Hopfiel model.}
\author{ M. Shcherbina$^\dagger$
 \qquad B. Tirozzi$^*$\\
$^\dagger$Institute for Low Temperature Physics, Ukr. Ac. Sci \\
$^*$Department of Physics, Rome University "La Sapienza"
 }
\date{}

\maketitle

\begin{abstract}
We find the free-energy in the thermodynamic limit of a one dimensional  XY model associated to
a system of N qubits. The coupling among the $\sigma_i^z$ is a long range two bodies random interaction. The randomness
in the couplings is the typical interaction  of the Hopfield model with $p$ patterns ($p<<N$), with the patterns being
a $p$ sequences of independent identically distributed (i.i.d.) random variables  assuming  values $\pm 1$ with
 probability
$1/2$.  We show also that in the case $p\le\alpha N$  the free-energy is asymptotically independent from the
 choice of the patterns, i.e. it is self-averaging.
The Hamiltonian is the one used by \cite{Glaser} in his experiment.

\end{abstract}

\section{Introduction and main results}\label{s:1}

The research on quantum computers has found many algorithms and most of them have been  implemented on real quantum systems,
like for example the Grover's algorithm  \cite{Grover}, or the quantum Fourier transform \cite{Quantum}. The
increase of the technology used in realizing quantum computers  \cite{Liquid}, has convinced a group of physicists
to try to implement a quantum neural network \cite{Glaser}. In their experiment they treat a system of two q-bits with
Hebbian interaction generated by one pattern with random entries $\pm 1$ :

\begin{equation}\label{I_H}
J_{ij}= \frac{1}{N} \sum_{\mu=1}^p \xi^\mu_i \xi^\mu_j,
\end{equation}
where $\{\xi^\mu_i\}_{i=1,\dots,N,\mu=1,\dots p}$ is a system of i.i.d. random variables, $\xi^\mu_i=\pm 1$ with probability
$\frac{1}{2}$.
Thus it is a model of the type introduced by Hopfield \cite{Hopfield} but the spins are quantum objects instead of the classical
Hopfield model. They used two q-bits of a NMR quantum computers
to which they applied the Quantum Adiabatic Computation (QAC)  \cite{QAC} for building the two q-bit state and they found
the retrieval states imposing some input Hamiltonian. The retrieval states were constructed with Quantum Annealing
\cite{SISSA} and they were the minima of the Hamiltonian . This is  the first example of implementation of a neural network
of the Hopfield type with quantum spins. There have been other proposal of quantum neural networks \cite{Ezhov} but we think that the system
built in \cite{Glaser} is more adequate to the function of retrieval and storage typical of the neural networks.
The problem is that the retrieval and storage in classical neural networks is connected with the self-averaging of the free-energy
\cite{ST:93}
and of the overlap parameters, i.e. the network stores and retrieves any kind of information which is translated in the formalism
as a retrieval for almost all patterns. In the classical model this result is obtained only in the thermodynamic limit
$N \to \infty$
and also in the quantum case there is a similar problem. The necessity of large values of $N$ arises from the application of
probability estimates and from the extension of the law of large numbers which is largely applied in the classical case.
We are aware of the fact that this limit is far beyond the actual possibilities of the experiments but nevertheless we think
that it is important to establish all the useful concepts and theoretical results. So in this work we treat the question of the
self-averaging of the free-energy of a one dimensional system of q-bits with Hebbian interaction and independent identically distributed random
patterns with values $\pm 1$ with equal probability. We compute also the free-energy and find the phase transition
and the  index for the critical temperature . The system is described by a Hamiltonian
of the type considered in \cite{Glaser}, so it is the XY model in the one dimensional case with Hebbian long range interaction, and
with the perturbation of the QAC term:

\begin{equation}
H=-\frac{1}{2}\sum_{i,j=1}^N J_{ij}\sigma_i^z\sigma_j^z-\sum_{i=1}^N h_i \sigma_i^z-\sum_{i=1}^Nd \sigma_i^x,
\label{H_H}\end{equation}
where
\begin{equation}
\sigma_i^z=\left(\begin{array}{cc}1&0\\0&-1\end{array}\right),\quad
\sigma_i^x= \left(\begin{array}{cc}0&1\\1&0\end{array}\right).\label{sigma}\end{equation}
This Hamiltonian has been considered also in other papers like for example \cite{Weindin} and
is the usual Hamiltonian for performing quantum measurements.
The term multiplying $\sigma_i^x$ is the one used in QAC for constructing the interesting quantum states, its presence is
fundamental for the phenomenology of the system. If $d=0$ this sytem is the usual Hopfield network which has been solved already
in \cite{Amit} with the non rigorous replica trick and in \cite{PST} using the rigorous cavity method. For $d <1$ and $d \to 1^-$ we find
a critical point for $h=0$ and $\beta \to \infty$ different from the usual Ising transitions.
The limiting value for the free energy and for
the overlap parameters has been obtained for  $p<<N$, i.e. in the limit $N\to \infty$ and $p/N\to 0$, so the capacity of the
network defined as usual as $\alpha= \frac{p}{N}$, goes to zero in the thermodynamic limit. The next step could be an
expansion of the asymptotic free-energy
for small $\alpha$ in a similar way as was done in \cite{ST:95}.

%
The main results of the paper are the following two theorems:
\begin{theorem}\label{t:0}
Consider the Hopfield model (\ref{H_H}) with   $h_i$ independent of
$\{\xi^{\mu}_j\}_{j\not=i}$ and such that $\mathbf{E}\{h_i^2\}\le C$. Then for any $\alpha\ge 0$, if
$p\to\infty$, $N\to\infty$, and $p/N\to\alpha$, then the free energy $f_N(\beta,H)$ is self averaging in the limit
$N\to\infty$
\begin{equation}
\mathbf{E}\big\{|f_N(\beta,H)-\mathbf{E}\{f_N(\beta,H)\}|^2\big\}\le C/N .\label{s-a}\end{equation}
Here and below we denote by $\mathbf{E}\{.\}$ the averaging with respect to all random parameters of the problem.
\end{theorem}
\begin{remark}\label{r:1} One can easily see  from the proof of Theorem \ref{t:0} that Bernoulli $\{\xi^\mu_i\}$
can be replaced by  i.i.d. $\{\xi^\mu_i\}$ with any distribution satisfying the conditions $\mathbf{E}\{\xi^\mu_i\}=0$,
$\mathbf{E}\{|\xi^\mu_i|^2\}=1$, and $\mathbf{E}\{|\xi^\mu_i|^{4+\varepsilon}\}\le C$. Moreover, the operators
$\sigma_i^z,\sigma_i^x$ of (\ref{sigma}) can be replaced by any bounded operators.
\end{remark}
\begin{theorem}\label{t:1}
Consider the Hopfield model (\ref{H_H}) with  $h_i=h\xi^1_i$. Then
 the mean free energy $E\{f_N(\beta,H)\}$ satisfies the inequality
\begin{equation}
|\mathbf{E}\{f_N(\beta,H)\}-\min_mf_0(m,h)|\le C \alpha_N^{1/3},\quad \alpha_N:=p/N \label{III.2.1}\end{equation} where
\begin{equation}
f_0(m,h)=-{1\over\beta}\log 2\cosh\beta ((m+h)^2+d^2)^{1/2}+\frac{m^2}{ 2} \label{f_0}\end{equation} is the free energy of the
Curie-Weiss model (the Hopfield model with only one pattern).
\end{theorem}
\begin{remark} Since the free energy is continuous function with respect to $h$, it follows from Theorem \ref{t:1} that
the statement of the theorem is valid also for $h=0$.
\end{remark}
\begin{remark}\label{r:3} From the proof of Theorem \ref{t:1} it will be seen that here also Bernoulli $\{\xi^\mu_i\}$
can be replaced by  i.i.d. $\{\xi^\mu_i\}$ with any distribution satisfying the same condition as in Remark \ref{r:1}.
But in this case the expression for $f_0(m,h)$ takes the form
\begin{equation}
f_0(m,h)=-{1\over\beta}\mathbf{E}\{\log 2\cosh\beta (\xi^{1}_1(m+h)^2+d^2)^{1/2}\}+\frac{m^2}{ 2}. \label{f_0.a}\end{equation}
 Moreover, the matrices
$\sigma_i^z,\sigma_i^x$ of (\ref{sigma}) can be replaced by spin matrices of any dimension.
But in this case  the expression for $f_0(m,h)$ will be different from (\ref{f_0.a}).

\end{remark}

\section{Proofs}\label{s:2}

The proofs of Theorems \ref{t:0}, \ref{t:1} are based on the method proposed in \cite{Bog}. We
start from the following general proposition:
\begin{proposition}\label{p:Bog} Let $H,H^1$ be any hermitian $2^N\times 2^N$ matrices, $H(t):=H+tH^1$
\[f_N(\beta,t)=-\frac{1}{\beta N}\log\mathrm{Tr\,}e^{-\beta H(t)},\quad
Z_N=\mathrm{Tr\,}e^{-\beta H(t)}\]
and $\{e_k\}_{k=1}^{2^N}$ are eigenvectors of $H(t)$, so that
\[H(t)e_k=E_ke_k,\quad H^1_{jk}:=(\overset\circ{H^1}e_k,e_j),\quad\overset\circ{H^1}:={H}^1-\la H^1\ra_{H(t)}.\]
Then
\begin{align}\label{pB.1}
-\frac{\partial^2}{\partial t^2}f_N(\beta,t)=&\frac{1}{ Z_N}\sum_{k,j=1}^{2^N}|H^1_{jk}|^2
\frac{e^{-\beta E_j}-e^{-\beta E_k}}{E_k-E_j}\ge 0, \end{align}
and the Bogolyubov inequality \cite{Bog} holds
\begin{equation}\label{pB.2}
\frac{1}{N}\la H^1\ra_{H(1)}\le f_N(\beta,1)-f_N(\beta,0)\le\frac{1}{N}\la H^1\ra_{H(0)}.
 \end{equation}
\end{proposition}
\textit{Proof.}
According to  the Duhamel formula we have
\begin{align*}
-\frac{\partial^2}{\partial t^2}f_N(\beta,t)=\frac{\beta}{N Z_N}\int_0^1\mathrm{Tr\,}
\Big(\overset\circ{H^1} e^{-\beta H(t)\tau}\overset\circ{H^1} e^{-\beta H(t)(1-\tau)}\Big)d\tau\\
=\frac{\beta }{N Z_N}\sum_{k,j=1}^{2^N}|H^1_{jk}|^2\int_0^1e^{-\beta E_k\tau-\beta E_j(1-\tau)}d\tau
=\frac{1}{ Z_N}\sum_{k,j=1}^{2^N}|H^1_{jk}|^2\frac{e^{-\beta E_j}-e^{-\beta E_k}}{E_k-E_j}.
\end{align*}
To prove (\ref{pB.2}) we observe that
\[-\frac{\partial^2}{\partial t^2}f(\beta,t)\ge 0\quad\Rightarrow\quad
\frac{\partial}{\partial t}f_N(\beta,1)\le\frac{\partial}{\partial t}f_N(\beta,t)\le\frac{\partial}{\partial t}f_N(\beta,0).\]
Integrating the last inequality with respect to $t$ from 0 to 1, we obtain (\ref{pB.2}). $\square$

\medskip

\textit{Proof of Theorem \ref{t:0}.} Denote $\mathbf{E}_{\le k}$ the averaging with respect to
$\{\xi^\mu_{i}\}_{1\le i\le k,1\le\mu\le p}$. Then,
according to the standard martingal method (see \cite{Dh-Co:68}), we have
\begin{align}\label{mart}\mathbf{Var}\{f_N(\beta,H)\}=\sum_{k=1}^n\mathbf{E}\{|\mathbf{E}_{\le k-1}\{f_N(\beta,H)\}-
\mathbf{E}_{\le k}\{f_N(\beta,H)\}|^2\}.\end{align}
Denote  $\mathbf{E}_{ k}$ the averaging with respect to
$\{\xi^\mu_{k}\}_{1\le\mu\le p}$ and $H^{(k)}:=H\Big|_{\xi^\mu_{k}=0,\mu=1,\dots,p}$. Then, using the Schwarz inequality, we obtain that
\begin{align*}&|\mathbf{E}_{\le k-1}\{f_N(\beta,H)\}-\mathbf{E}_{\le k}\{f_N(\beta,H)\}|^2=
|\mathbf{E}_{\le k-1}\{f_N(\beta,H)-\mathbf{E}_{ k}\{f_N(\beta,H)\}|^2\\ &\le
\mathbf{E}_{\le k-1}\{|f_N(\beta,H)-E_{ k}\{f_N(\beta,H)\}|^2\}\le\mathbf{E}_{\le k-1}\{|f_N(\beta,H)-f_N(\beta,H^{(k)})|^2\}.
\end{align*}
Hence
\begin{equation}\label{b_v.2}
    \mathbf{Var}\{f_N(\beta,H)\}\le\sum_{k=1}^n\mathbf{E}\{|f_N(\beta,H)-f_N(\beta,H^{(k)})|^2\}.
\end{equation}
By the Bogolyubov inequality
\begin{align*}\la\Delta h_k\sigma^z_k\ra_{H}+\frac{1}{N}\Big\la\sum_{j=1}^NJ_{kj}\sigma^z_k\sigma^z_j\Big\ra_{H}&\le
f_N(\beta,H)-f_N(\beta,H^{(k)})\\
&\le\la\Delta h_k\sigma^z_k\ra_{H^{(k)}}+\frac{1}{N}\Big\la\sum_{j=1}^NJ_{kj}\sigma^z_k\sigma^z_j\Big\ra_{H^{(k)}},\end{align*}
where $\Delta h_k=h_k-h_k\Big|_{\xi^\mu_{k}=0,\mu=1,\dots,p}$. Hence
\begin{align}\label{b_v.3}\mathbf{E}\{|f_N(\beta,H)-f_N(\beta,H^{(k)})|^2\}\le&
\mathbf{E}\Big\{\Big\la\frac{1}{N}\sum_{j=1}^NJ_{kj}\sigma^z_k\sigma^z_j\Big\ra_{H^{(k)}}^2\Big\}\\&+
\mathbf{E}\Big\{\Big\la\frac{1}{N}\sum_{j=1}^NJ_{kj}\sigma^z_k\sigma^z_j\Big\ra_{H}^2\Big\}
+4\mathbf{E}\{h_k^2\}.\notag\end{align}
Since $H^{(k)}$ does not depend on $\{\xi^\mu_{k}\}_{1\le\mu\le p}$, averaging with respect to
$\{\xi^\mu_{k}\}_{1\le\mu\le p}$ and using that
\[\mathbf{E}\{J_{ik}J_{jk}\}=N^{-1}J_{ij},\]
 we get
\[\sum_{k=1}^N\mathbf{E}\Big\{\Big\la\frac{1}{N}\sum_{j=1}^NJ_{kj}\sigma^z_k\sigma^z_j\Big\ra_{H^{(k)}}^2\Big\}=
\sum_{k=1}^N\mathbf{E}\Big\{\frac{1}{N^2}\sum_{i,j=1}^NJ_{ij}\la\sigma^z_k\sigma^z_i\ra_{H^{(k)}}
\la\sigma^z_k\sigma^z_j\ra_{H^{(k)}}\Big\}\le \frac{\mathbf{E}\{||J||\}}{N}.\]
For the second term in the r.h.s. of (\ref{b_v.3}) after summation with respect to $k$ we get
\begin{align*}
&\sum_{k=1}^N\mathbf{E}\Big\{\frac{1}{N^2}\sum_{i,j=1}^NJ_{ki}J_{kj}
\la\sigma^z_k\sigma^z_i\ra_{H}\la\sigma^z_k\sigma^z_j\ra_{H}\Big\}\le
\sum_{k=1}^N\mathbf{E}\Big\{\frac{1}{N^2}\sum_{i,j=1}^NJ_{ki}J_{kj}
\la(\sigma^z_k)^2\sigma^z_i\sigma^z_j\ra_{H}\Big\}\\
&=\mathbf{E}\Big\{\frac{1}{N^2}\sum_{i,j,k=1}^N(J^2)_{ij}
\la\sigma^z_i\sigma^z_j\ra_{H}\Big\}\le \frac{\mathbf{E}\{||J^2||\}}{N}.
\end{align*}
Since it is well known that (see, e.g. \cite{ST:93})
\begin{equation}\label{||J||}
\mathbf{E}\{||J^2||\}\le C,
\end{equation}
the last two bounds combined with (\ref{b_v.2}) and (\ref{b_v.3}) prove Theorem \ref{t:0}. $\square$

\medskip

\textit{Proof of Theorem \ref{t:1}}. To prove Theorem \ref{t:1} let us introduce some additional Gaussian field to the Hamiltonian $H$
\begin{equation}
 H(\overline\gamma)=H+\sqrt{N}\sum_{\mu=1}^p\gamma_\mu m^\mu,
\label{H(g)}\end{equation}
where
\begin{equation}
 m^\mu:=\frac{1}{ N}\sum^N_{i=1}\xi^\mu_i \sigma_i^z,
\label{mm}\end{equation}
are so-called overlaps parameters and $\{\gamma_\mu\}_{\mu=1}^p$ are independent of $\{\xi^\mu_j\}$
and of each other Gaussian random variables with zero mean and variance 1.
Using the Bogolyubov inequality (\ref{pB.2}) and then the Schwarz inequality and (\ref{||J||}), we get
\begin{align}\notag
 0&\le\mathbf{E}\{f_N(\beta,H(\overline{\gamma}))\}-\mathbf{E}\{f_N(\beta,H)\}\\
 &\le\frac{1}{\sqrt{N}}\mathbf{E}\Big\{\sum_{\mu=1}^p\gamma_\mu \la m^\mu\ra_{H(\overline\gamma)}\Big\}
 \le\alpha_N^{1/2}\mathbf{E}^{1/2}\Big\{
 \sum_{\mu=1}^p\la m^\mu\ra_{H(\overline\gamma)}^2\Big\}\notag\\&= \frac{\alpha_N^{1/2}}{N}\mathbf{E}^{1/2}
 \Big\{\sum_{i,j=1}^NJ_{ij}\la\sigma_i^z\ra_{H(\overline\gamma)}\la
 \sigma_i^z\ra_{H(\overline\gamma)}\Big\}\le C\alpha_N^{1/2}.
\label{in.1}\end{align}
Consider also the "approximate" Hamiltonian of the form
\begin{align}\notag
H_a(\overline \gamma,\overline c)=&H(\overline \gamma)+\frac{N}{2}\sum_{\mu=1}^p(m^\mu-c^\mu)^2\\=&
-N\sum_{\mu=1}^p m^\mu c^\mu-\sum_{i=1}^N h_i \sigma_i^z-\sum_{i=1}^Nd \sigma_i^x+\frac{N}{2}\sum_{\mu=1}^p(c^\mu)^2
+\sqrt{N}\sum_{\mu=1}^p\gamma_\mu m^\mu.
\label{H_a}\end{align}
Note that similarly to (\ref{in.1}) we have uniformly in $\overline{c}\in\mathbb{R}^p$
\begin{align}
\Big| \mathbf{E}\{f_N(\beta,H_a(\overline{\gamma},\overline{c}))\}-\mathbf{E}\{f_N(\beta,H_a(0,\overline{c}))\}\Big|
 \le \alpha_N^{1/2}C.
\label{in.2}\end{align}
By the Bogolyubov inequality (\ref{pB.2}) for any $\overline{c}\in\mathbb{R}^p$
\begin{equation}
 0\le f_N(\beta,H_a(\overline{\gamma},\overline{c}))-f_N(\beta,H(\overline{\gamma}))
 \le\frac{1}{2}\sum_{\mu=1}^p\la(m^\mu-c^\mu)^2\ra_{H(\overline{\gamma})}.
\label{Bog}\end{equation}

Hence
\begin{equation}
 0\le \min_{\mathbf{c}\in\mathbb{R}^p}f_N(\beta,H_a(\overline{\gamma},\overline{c}))-f_N(\beta,H(\overline{\gamma}))
 \le\frac{1}{2}\sum_{\mu=1}^p
 \la(\overset{\circ}m^\mu)^2\ra_{H(\overline{\gamma})}.
\label{Bog1}\end{equation}
Here and below we denote
\begin{equation}
 \overset{\circ}m^\mu:=m^\mu-\la m^\mu\ra_{H(\overline{\gamma})}.
\label{ov_m}\end{equation}
Let $\{\overline{e}_k\}_{k=1}^{2^N}$ be the basis in which $H(\overline\gamma)$ is diagonal and
\[H(\overline\gamma)\overline{e}_k=E_k\overline{e}_k,\quad M^\mu_{jk}=(\overset{\circ}m^\mu\overline{e}_k,\overline{e}_j).\]
It is easy to see that
\[\sum_{\mu=1}^p\la(m^\mu-\la m^\mu\ra)^2\ra_H=
\frac{1}{ Z_N}\sum_{\mu=1}^p\sum_{k,j=1}^{2^N}|M^\mu_{jk}|^2\frac{1}{2}(e^{-\beta E_j}+e^{-\beta E_k}).\]

Using the inequality
\[\cosh x\le\frac{\sinh x}{x}+|\sinh x|,\]
we can write
\begin{align*}
\sum_{\mu=1}^p\la(m^\mu-\la m^\mu\ra)^2\ra_H&=
\frac{1}{ Z_N}\sum_{\mu=1}^p\sum_{k,j=1}^{2^N}|M^\mu_{jk}|^2\frac{1}{2}(e^{-\beta E_j}+e^{-\beta E_k})\\&\le
\frac{1}{ Z_N}\sum_{\mu=1}^p\sum_{k,j=1}^{2^N}|M^\mu_{jk}|^2\Big(\frac{e^{-\beta E_j}-e^{-\beta E_k}}{E_k-E_j}
+\frac{1}{2}|e^{-\beta E_j}-e^{-\beta E_k}|\Big)\\&=
-\sum_{\mu=1}^p\frac{\partial^2}{\partial\gamma_\mu^2}f_N(\beta,H(\overline{\gamma}))+\Sigma,
\end{align*}
where we have used that according to (\ref{pB.1})
\[\frac{1}{ Z_N}\sum_{k,j=1}^{2^N}|M^\mu_{jk}|^2\frac{e^{-\beta E_j}-e^{-\beta E_k}}{E_k-E_j}=
-\frac{\partial^2}{\partial\gamma_\mu^2}f_N(\beta,H(\overline{\gamma})).\]
To estimate $\Sigma$, we use the Holder inequality, which yields
\begin{align*}
\Sigma:&=\frac{1}{ 2Z_N}\sum_{\mu=1}^p\sum_{k,j=1}^{2^N}|M^\mu_{jk}|^2|e^{-\beta E_j}-e^{-\beta E_k}|\\&\le
\frac{1}{ 2Z_N}\sum_{\mu=1}^p\sum_{k,j=1}^{2^N}|M^\mu_{jk}|^2\frac{|e^{-\beta E_j}-e^{-\beta E_k}|}{|E_k-E_j|}|E_k-E_j|
 \end{align*}
\begin{align*}
&\le\Big(\frac{1}{ 2Z_N}\sum_{\mu=1}^p\sum_{k,j=1}^{2^N}|M^\mu_{jk}|^2\frac{|e^{-\beta E_j}-e^{-\beta E_k}|}{|E_k-E_j|}\Big)^{2/3}
\\&\times\Big(\frac{1}{ 2Z_N}\sum_{\mu=1}^p\sum_{k,j=1}^{2^N}|M^\mu_{jk}|^2(e^{-\beta E_j}+e^{-\beta E_k})|E_k-E_j|^2\Big)^{1/3}\\
&=\Big(-\sum_{\mu=1}^p\frac{\partial^2}{\partial\gamma_\mu^2}f_N(\beta,H(\overline{\gamma}))\Big)^{2/3}\Sigma_1^{1/3}.
\end{align*}
But it is easy to see that
\begin{align*}
\Sigma_1:&=\frac{1}{ 2Z_N}\sum_{\mu=1}^p\sum_{k,j=1}^{2^N}|M^\mu_{jk}|^2(e^{-\beta E_j}+e^{-\beta E_k})|E_k-E_j|^2
=-\sum_{\mu=1}^p\la[m^\mu,H]^2\ra_H\\
&=\sum_{\mu=1}^p\Big\la\Big(\frac{1}{N}\sum_{j=1}^N\xi^\mu_j
[\sigma^z_j,d\sigma^x_j]\Big)^2\Big\ra_H
=\frac{4d^2}{N}\sum_{i,j=1}^NJ_{ij}\la\sigma^y_i\sigma^y_j\ra\le 4d^2||J||.
\end{align*}
On the other hand, averaging with respect to the Gaussian variables $\gamma_\mu$, we obtain similarly to (\ref{in.1})
\begin{align*}&\mathbf{E}\Big\{-\sum_{\mu=1}^p\frac{\partial^2}{\partial\gamma_\mu^2}f_N(\beta,H(\overline{\gamma})\Big\}=
\mathbf{E}\Big\{-\sum_{\mu=1}^p\gamma_\mu\frac{\partial}{\partial\gamma_\mu}f_N(\beta,H(\overline{\gamma})\Big\}\\
&=-\mathbf{E}\Big\{\frac{1}{\sqrt N}\sum_{\mu=1}^p\gamma_\mu \la m^\mu\ra_H\Big\}\le\mathbf{E}^{1/2}\Big\{\frac{1}{N}
\sum_{\mu=1}^p\gamma_\mu^2\Big\}\mathbf{E}^{1/2}\Big\{\frac{1}{N}\sum_{i,j=1}^NJ_{ij}\la\sigma^z_i\ra\la\sigma^z_j\ra\Big\}\le
C\alpha_n^{1/2}.
\end{align*}
The above inequalities combined with (\ref{Bog1}), (\ref{in.1}) and (\ref{in.2}) yield
\[-C\alpha_n^{1/2}\le\mathbf{E}\{\min_{\overline c\in\mathbb{R}^p}f_N(\beta,H^a(0,\overline c))\}-
\mathbf{E}\{f_N(\beta,H)\}\le C\alpha_N^{1/3}.\]
In view of this bound it suffices to find
$\mathbf{E}\{\min_{\overline c\in\mathbb{R}^p}f(H^a(0, \overline c))\}$. By using the convexity of the function
$-\log2\cosh\beta\sqrt{x}$ in $x$ for $x>0$, we get for
$C_i=\xi^1_i h+\sum_{\mu}\xi^\mu_i c^\mu$
\begin{align}\notag
f(H^a(0,\overline c))=&
-\frac{1}{\beta N}\sum_{i=1}^N\log 2\cosh\beta (C_i^2+d^2)^{1/2}
+\frac{1}{ 2}\sum_{\mu=1}^p(c^\mu)^2\\\ge&
-\frac{1}{ \beta}\log 2\cosh\beta\Big(\frac{1}{N}\sum_{i=1}^NC_i^2+d^2\Big)^{1/2}+
\frac{1}{ 2}\sum_{\mu=1}^p(c^\mu)^2\notag\\=&
-\frac{1}{ \beta}\log 2\cosh\beta\Big((c^1+h)^2+\sum_{\mu=2}^p(c^\mu)^2+d^2\notag\\&
+\big( A(\overline c+h\bar e^1),(\overline c+h\bar e^1)\big)\Big)^{1/2}
+\frac{1}{2}\sum_{\mu=1}^p(c^\mu)^2\notag\\\ge&
-\frac{1}{ \beta}\log 2\cosh\beta\Big((c^1+h)^2+\sum_{\mu=2}^p(c^\mu)^2+d^2\Big)^{1/2}+
\frac{1}{2}\sum_{\mu=1}^p(c^\mu)^2\label{in_min}\\&-\frac{|| A||}{2}(h^2+(\overline c,\overline c)).
\notag\end{align}
Here  $\bar e^1=(1,0, \dots,0)\in\mathbb{R}^p$,
 and the matrix $A$ is defined  as
\begin{equation}
A^{\mu\nu}={1\over N}( 1- \delta_{\mu\nu}) (\mathbf{\xi}^\mu, \mathbf{\xi}^\nu).
\label{A}\end{equation}
It is  known (see e.g. \cite{ST:93}) that
\begin{equation}
\mathbf{E}\{||A||^2\}\le C\alpha_N.
\label{||A||}\end{equation}
Moreover, it is easy to see that if $\overline c$ is a minimum point, then
\[ c^\mu=\la m^\mu\ra_{H_a} \quad \Rightarrow\quad
\sum_{\mu=2}^p(c^\mu)^2=\frac{1}{N}\sum J_{ij}\la\sigma^z_i\ra_{H_a}\la\sigma^z_j\ra_{H_a}\le
||J||.\]
Hence (\ref{in_min}) and (\ref{||J||}) imply
\begin{align}\notag
&\mathbf{E}\{\min_{\overline c\in\mathbb{R}^p} f(H^a(0,\overline c))\}\\
&\ge\min_{\overline c\in\mathbb{R}^p}
\Big\{-\frac{1}{ \beta}\log 2\cosh\beta\Big((c^1+h)^2+\sum_{\mu=2}^p(c^\mu)^2+d^2\Big)^{1/2}+
\frac{1}{2}\sum_{\mu=1}^p(c^\mu)^2\Big\}-C\sqrt{\alpha_N}\notag\\
&\ge\min_{r\ge 0,0\le\varphi\le 2\pi}\Big\{-\frac{1}{ \beta}\log 2\cosh\beta\Big((r\sin\varphi+h)^2
+r^2\cos^2\varphi+d^2\Big)^{1/2}+\frac{r^2}{2}\Big\}-C\sqrt{\alpha_N}\notag\\
&\ge\min_{m\ge 0}\Big\{-\frac{1}{ \beta}\log 2\cosh\beta\Big((m+h)^2+d^2\Big)^{1/2}+\frac{m^2}{2}\Big\}-C\sqrt{\alpha_N}.
\label{}\end{align} $\square$

Let us now discuss briefly  the equation for the point in which the r.h.s. of (\ref{f_0}) attains its minimum.
It is easy to see that it has the form
\begin{equation}\label{eq1}
    m=\frac{(m+h)\tanh\beta\big((m+h)^2+d^2\big)^{1/2}}{\big((m+h)^2+d^2\big)^{1/2}}.
\end{equation}
For $h=0$ it takes the form
\begin{equation}\label{eq2}
    m=\frac{m\tanh\beta\big(m^2+d^2\big)^{1/2}}{\big(m^2+d^2\big)^{1/2}}.
\end{equation}
It is evident that it always has a solution $m=0$.  To find another solution we should study the equation
\begin{equation}\label{eq3}
    \big(m^2+d^2\big)^{1/2}=\tanh\beta\big(m^2+d^2\big)^{1/2}.
\end{equation}
This equation for $d>1$ has no solutions because the r.h.s. is less than 1 and the l.h.s. is more than 1.
For $\beta<1$ the equation also has no solutions, since it is well known that the equation $\tanh\beta x=x$ has
no solutions except $x=0$ for $\beta<1$.

For $d<1$ there is a critical point $\beta (d)$ such that for $\beta>\beta (d)$ (\ref{eq2}) has the unique solution
$m=0$ and $\beta>\beta (d)$ there is also non zero solution of for  (\ref{eq2}).  This critical
value  $\beta (d)$ is a  solution of the equation
\begin{equation}\label{eq4}
   d=\tanh\beta d.
\end{equation}
It is easy to see that $\beta(0)=1$ and $\beta(d)\to\infty$ as $ d\to 1$ ($d<1$). One can see also that $\beta'(d)\ge 0$ since
it follows from (\ref{eq4})
\[\beta'(d)=d^{-1}\cosh^2\beta d\Big(1-\frac{\beta}{\cosh^2\beta d}\Big)\]
and the r.h.s. here is positive, since $\dfrac{\beta}{\cosh^2\beta d}$ is the derivative of the r.h.s. of (\ref{eq4})
with respect to $d$, and at the solution point of  (\ref{eq4}) this derivative is less than 1.
Moreover, since $\tanh(\beta d)\sim 1- e^{-2\beta d}$ for $\beta d\to\infty$, we have
from (\ref{eq4}) that
\[ e^{2\beta d}\sim (1-d)^{-1}\quad \Rightarrow\quad \beta(d)\sim\frac{1}{2}\log (1-d)^{-1},\quad d\to 1.\]

\medskip

\textbf{Acknowledgements}. M.S. is grateful the Italian National  Group of Mathematical Physics  for the financial support during her stay in Italy.

\end{document}